\documentclass[11pt,twoside]{article}

\usepackage{asp2006}
\usepackage{epsf}
\usepackage{lscape}

\begin{document}
\title{Habitability of Planets in Binaries}  
\author{Nader Haghighipour}   
\affil{Institute for Astronomy and NASA Astrobiology Institute,
University of Hawaii-Manoa}   

\begin{abstract} 
A survey of currently known extrasolar planets indicates that close to 
20\% of their hosting stars are members of binary systems. While the 
majority of these binaries are wide (i.e., with separations between 250 
and 6500 AU), the detection of Jovian-type planets in the three binaries of
$\gamma$ Cephei (separation of 18.5 AU), GL 86 (separation of 21 AU),
and HD 41004 (separation of 23 AU) 
have brought to the forefront questions on the formation of giant planets 
and the possibility of the existence of smaller bodies in moderately 
close binary star systems. This paper
discusses the late stage of the formation of habitable planets in 
binary systems that host
Jovian-type bodies, and reviews the effects of the binary companion 
on the formation of Earth-like planets in the system's habitable zone. 
The results of a large survey of the parameter-space of 
binary-planetary systems in search of regions where habitable planets 
can form and have long-term stable orbits are also presented. 
\end{abstract}

\section{Introduction}

Among the currently known planet-hosting stars, approximately 20\% 
are members of binary systems \citep{Hagh06}. Although 
to observers, the existence of such {\it binary-planetary} 
systems is not unexpected \footnote{Observations of star-forming 
regions have indicated that a large
fraction of main and pre-main sequence stars are formed in dual
or multistar environments \citep{Abt79,Duq91,Math94,Math00,White01}. 
There is also substantial evidence
on the existence of planet-forming disks around stellar components 
of binary systems \citep{Math94,Akeson98,Rodriguez98,White99,
Silbert00,Math00}.}, to theorists, such 
{\it extreme} planetary environments pose major challenges
to theories of planet formation. 
While observations of systems 
such as L1551 \citep{Rodriguez98} indicate that planet-forming 
circumstellar disks, with
masses similar to the mass of the primordial nebula of our
solar system, exist around the components of binaries (implying 
that planet-formation in dual-star systems can begin and proceed in
the same fashion as around our Sun), simulations of the formation
of giant planets around stars of a binary yield mixed results. For instance,
as shown by \citet{Nelson00}, giant planets cannot form in binary systems with
separations of $\sim 50$ AU through disk instability or core 
accretion mechanisms. However, as shown by 
\citet{Boss06} and by \citet{Mayer07}, disk instability 
can indeed form giant planets in binary systems, and 
as indicated by \citet{Thebault04},
the core-accretion mechanism is also capable of
forming planets in dual-star systems [see \citet{Hagh07b}
for a comprehensive review].

The fact that giant planets exist in binary systems 
implies that planet formation in dual-star environments
is robust. One important concern with such systems
is, then, whether they can also form and harbor
habitable planets. In this paper,
we study habitable planet formation in binary systems that are 
moderately close (i.e., separation smaller then 50 AU), and also
harbor giant planets. Among the currently known binary-planetary
systems only three are of this kind:
GL 86 \citep{Els01}, $\gamma$ Cephei \citep{Hatzes03}, 
and HD 41004 \citep{Zucker04,Ragh06}. 
We are in particular interested to understand
how the dynamics of the secondary star affects the late stage of the
formation of terrestrial-class planets in the habitable zone
of the primary and the delivery of water
to its habitable planets.

\section {Numerical Simulations}

To study the late stage of habitable planet formation,
the collisional growth of a few hundred Moon- to Mars-sized objects
(planetary embryos) has to be simulated. In a recent article
\citep{Hagh07}, we carried out such simulations for a binary-planetary
system with a solar-type star as its primary, a Jupiter-sized
object at 5 AU as its planetary companion, and a 0.5-1.5 solar-masses
star as its secondary.
In these simulations, we assumed that planetesimal
formation has been efficient and has resulted in the formation of a disk of
approximately 115 planetary embryos with masses ranging from 0.01 to 0.1 
Earth-masses. We randomly distributed these objects between 0.5 AU and 4 AU 
by 3 to 6 mutual Hill radii, and considered the increase in their 
masses with their semimajor axes $(a)$ and the number of their 
mutual Hill radii $(\Delta)$ follow \citet{Raymond04}, and be given by 
${a^{3/4}}{\Delta^{3/2}}$. The surface density of our disk model, 
normalized to a density of 8.2 g/cm$^2$ at 1 AU, was assumed to follow
an $r^{-1.5}$ profile, where
$r$ is the radial distance from the primary star. We also 
assumed that the water to mass ratios  
of embryos followed the current distribution of water in
primitive asteroids of the asteroid belt \citep{Abe00}. That is,
embryos inside 2 AU were taken to be dry, the ones
between 2 and 2.5 AU were considered to contain 1\% water, and those
beyond 2.5 AU were assumed to have water to mass ratios of 5\%
\citep{Raymond04}.
\begin{figure}
\vskip -1.2in
\plotone {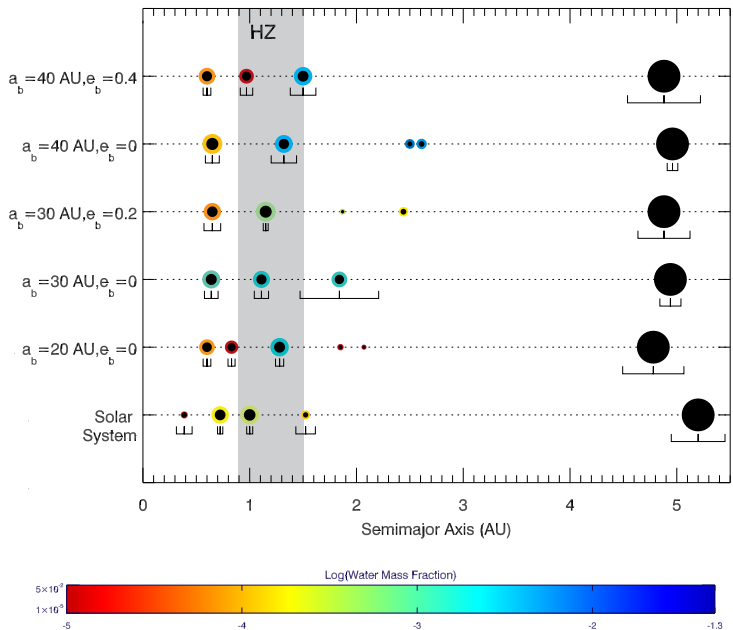}
\vskip 0.2in
\caption{Formation of water-bearing terrestrial-class planets around
the primary of a binary-planetary system with solar-mass stars. 
The big black circles 
represent the Jupiter-like planet of the system. Figure from \citet{Hagh07}.}
\label{fig1}
\end{figure}

We simulated the collisional growth of the planetary embryos of our 
binary-planetary system for 100 Myr,
and for different values of the semimajor axis $(a_b)$, orbital eccentricity
$(e_b)$, and mass of the secondary star.  
We followed \citet{Kasting93}, and considered a conservative
habitable zone for the primary of our system
at a distance
between 0.9 AU and 1.5 AU from this star. Figure 1 shows the results of some of
our simulations for an equal-mass binary with solar-type stars. 
As shown here, terrestrial-class planets with substantial
amounts of water can form in the habitable zone of the primary star.

An important result obtained from our simulations was the relation
between the binary perihelion $(q_b)$ and the location 
of the outermost terrestrial planet $(a_{\rm out})$. 
The left graph of figure 2 shows this for different simulations. 
As shown here, in simulations with no giant planet, similar 
to \citet{Quintana07},
terrestrial planet formation favors regions interior to $0.19 {q_b}$. 
Given the location of the inner edge of the habitable zone 
(i.e., 0.9 AU), our simulations indicated that, in a binary-planetary system
with a Sun-like primary, a stellar companion with a perihelion distance 
smaller than 0.9/0.19 = 4.7 AU$\sim$5 AU may not allow Earth-like
planets to form in the system's habitable zone. 
In simulations with giant planets, on the other hand, figure 2
shows that terrestrial planets form closer-in. 
The ratio $a_{out}/q_b$ in these 
systems varies between 0.06 and 0.13. 

A detailed analysis of the results of our simulations
indicate that the systems, in which water-bearing planets were formed
in their habitable zones,
have relatively large perihelia. The right graph of figure 2 shows this for 
simulations in a binary with equal-mass Sun-like stars.
The circles in this figure correspond to systems in which
simulations resulted in the formation of habitable planets. 
The numbers on the top of the circles represent
the mean eccentricity of the giant planet during the simulation. 
For comparison, the systems in which
the giant planet is unstable have also been marked.
Since at the beginning of each simulation,
the orbit of the giant planet was considered to be
circular, a non-zero eccentricity is indicative of the interaction 
of this body with the secondary star. As shown here, Earth-like objects 
are formed in systems where the average eccentricity of the giant 
planet is small. That is, in systems where the interaction 
between the giant planet and the secondary star has been weak. 
That implies, habitable planet formation is more favorable in
binaries with moderate to large perihelia, and with giant
planets on low eccentricity orbits.

\begin{figure}
\plottwo {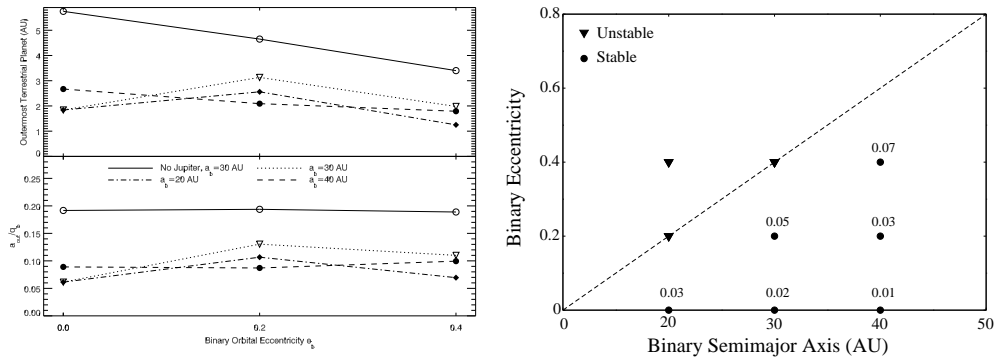}{f3.eps}
\vskip 0.2in
\caption{The graph on the left shows the relation between the perihelion
of an equal-mass binary and the location of its 
outermost terrestrial planet. 
The graph on the right shows the region of the $({e_b},{a_b})$ space
for a habitable binary-planetary system. Figures from \citet{Hagh07}.}
\label{fig2}
\end{figure}

\acknowledgements
Support by the NASA Astrobiology Institute under Cooperative Agreement 
NNA04CC08A with the Institute for Astronomy at the University of
Hawaii-Manoa is acknowledged.

\end{document}